\documentclass[aip,sd, amsmath,amssymb, reprint,onecolumn]{revtex4-1}
\usepackage[british,UKenglish,USenglish,english,american]{babel}
\usepackage{graphicx}
\usepackage{dcolumn}
\usepackage{bm}
\usepackage[pdftex]{color}
\usepackage{tikz}        
\usepackage{tikz}  
\usetikzlibrary{decorations.pathmorphing}
\usetikzlibrary{patterns}
\usepackage{xcolor} 
\usepackage{pgfplots}
\usetikzlibrary{shapes.geometric}
\usetikzlibrary{calc,3d}
\newcommand {\bydef}{\,\raise.07485ex\hbox{:}\kern-.025em\hbox{=}\,}
\newcommand {\Sc}  {\mathcal{S}}

\newcommand{\Jd} {J}

\newcommand{\lmg} {\gamma}
\begin{document}

\title{Modeling liquid migration in active swollen gel spheres}
%
\author{Michele Curatolo}%
\email[]{michele.curatolo@uniroma1.it}
%
\author{Paola Nardinocchi}%
\email[]{paola.nardinocchi@uniroma1.it}
\affiliation{Sapienza, Universit\`a di Roma, Roma, Italy}
\author{Luciano Teresi}%
\email[]{luciano.teresi@uniroma1.it}
\affiliation{Universit\`a di Roma Tre, Roma, Italy}
\date{\today}

\begin{abstract}
Liquid migration in active soft solids is a very common phenomenon in Nature at different scales: from cells to leaves. It can be caused by  mechanical as well as  chemical actions. The work focuses on the migration of liquid provoked by remodeling processes in an active impermeable gel sphere. Within this context, we present a consistent mathematical theory capable to gain a deep understanding of  the phenomenon in both steady and transient conditions.
\end{abstract}

\maketitle


%
%
Active soft matter is the key constituent of living matter: its striking behavior is the capability of exploiting chemical energy to produce mechanical work, and thus move, morph  and remodel. Cells are a prototypical example: their mechanical behavior  is controlled by a network of crosslinked filaments which respond to energy-transducing molecular motors\cite{Moeenda:2013}; the system is ``out of thermodynamic equilibrium'' and its functioning is based on the conversion of chemical power into mechanical power.

\begin{figure}[tb]
\includegraphics[width=0.5\textwidth]{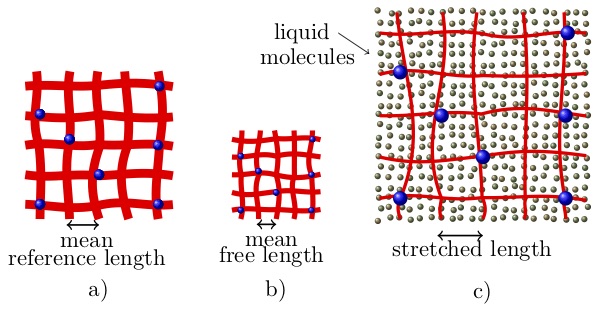}
\caption{\label{fig:1}  Three states of an active gel. a) Dry-reference state of the polymer network (red) with crosslinks (blue dots). b) Dry-remodeled (contracted) network: mean free-length is reduced and polymer chains are still un-stretched whereas the crosslink density changed.  c) Swollen state: liquid molecules (brown dots) swell the dry-contracted network.}
\end{figure}

\begin{figure}[h]
\includegraphics[width=0.75\textwidth]{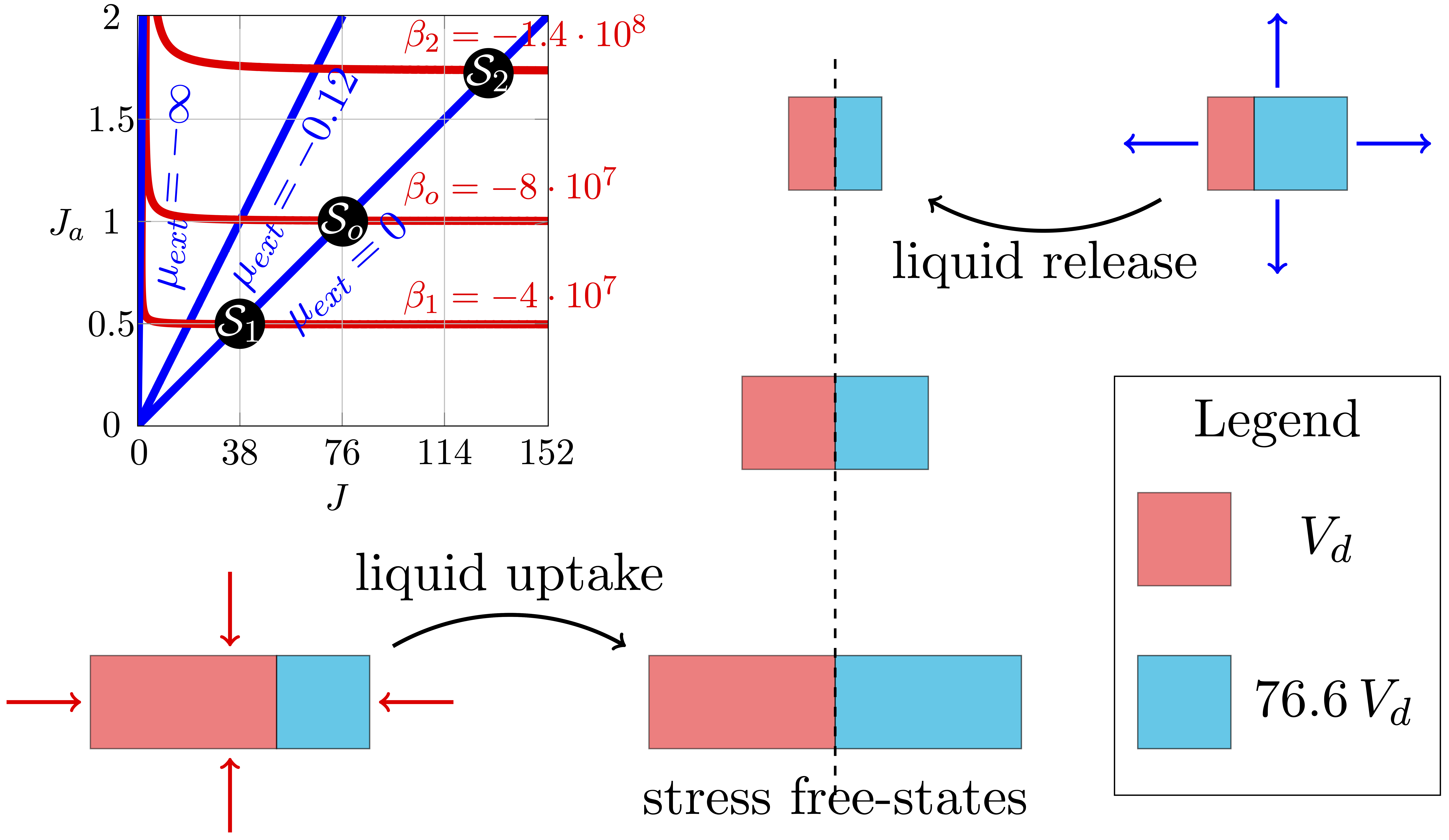}
\caption{\label{fig:2} Activation-induced liquid release and uptake. 
Top left) State space $(J, J_a)$; the intersections between iso-$\mu_{\rm{ext}}$ (blue) and iso-$\beta$ (red) represent
states that are both stress-free and steady.  The three highlighted states $\Sc_i$ refer at our example: starting from $\Sc_o$,
it is possible to halve the volume and get $\Sc_1$ by increasing $\beta$; analogously, by decreasing $\beta$
the volume can double to $\Sc_2$. All the states along the blue line $\mu_{\rm{ext}}$ have the same value $J_{eo}=77.6$.
Center) The three states $\Sc_i$ are shown in a stack, with cartoons representing polymer in red and liquid in cyan
(volumes not in scale); the middle cartoon represents $\Sc_o$, with $J_a=1$ and $J=J_{eo}\,J_a=77.6$; 
above we have $\Sc_1$, with $J_a=0.5$ and $J=J_{eo}\,J_a=38.8$; below we have $\Sc_2$, with $J_a=2$ and $J=J_{eo}\,J_a=155.2$.
Top right) Contracted swollen polymer with $J_a=0.5$ and same liquid content as $\Sc_o$: as $J_e>J_{eo}$, this state is under tension
and liquid must be released.
Bottom left) Expanded swollen polymer with $J_a=2$ and same liquid content as $\Sc_o$: as $J_e<J_{eo}$, it is under compression
and liquid must be absorbed.}
\end{figure}

In the more recent years, the study of this kind of active system has been gaining  an important role in both physics, as we need new experiments, and in mathematics, 
as we need new models\cite{Prost:2015}. 
The present study focuses on a theory of active gels on the macroscale\cite{Armstrong:2016,Curatolo:2017_M,Bacca:2019,Curatolo:2019} and aims to understand the mechanism under liquid redistribution which can be observed in active gels and in other macroscopic living systems where it delivers stresses affecting cell growth\cite{Kroeger:2011,Beauzamy:2014,Sahaf:2016,Zhang:2020}.\\  
We use the perspective of continuum physics: an active gel is considered as a biphasic material consisting of an elastic polymer network bathed in an interstitial liquid. Activation produces a change in the mean free-lengths of the polymer chains: at the macroscale,
this phenomenon is viewed as a change of the natural state of the 
network and named \textit{material remodeling}. Liquid redistribution, network deformation and material remodeling are strongly coupled and affect one each other. They determine stresses within the gel which can alter material remodeling and gel shape and may drive further remodeling\cite{Kroeger:2011,Xu:2020,Fei:2020}.\\ 
The state variables of the model,  which is based on an augmented version of the classic Flory-Rehner thermodynamics for stress diffusion\cite{FR1,FR2} put within the framework of material remodeling,\cite{Rodriguez:1994,DiCarlo:2002,Gurtin:2000} include  remodeling variables: they describe the change in size and shape of volume elements due to the changes of the mean free-lengths 
of the polymer chains\cite{Curatolo:2017_M,Curatolo:2019} (see figure \ref{fig:1}) As a secondary even if relevant effect, remodeling changes polymer network crosslink density too\cite{Bacca:2019}. Lastly, active gels allow to discuss the interaction between liquid migration and stress generation which are so important in the remodeling of plant cells\cite{Kroeger:2011}.\\
%
%
%
%
Let us start from  passive gels, swollen at  equilibrium. A gel placed in a liquid changes its volume by absorbing or desorbing the liquid and eventually reaches thermodynamic equilibrium. Whenever constraint-free and under no loads, the equilibrium state of a homogeneous gel is stress-free, and its swelling is uniform. For example, a  homogeneous gel sphere $\Sc_d$ with dry-reference radius $r_d$ (and dry-volume $V_d$) placed in a bath of chemical potential $\mu_o$ grows to get thermodynamical equilibrium and reaches  a stress--free swollen state $\Sc_o$ with  radius $(J^{1/3})r_d$. The change in volume $J$ is known from the chemical equilibrium equation which holds within the classic Flory-Rehner model of stress-diffusion
\begin{equation}\label{steady1}
R\,T\Bigl(\textrm{log}\frac{J-1}{J} + \frac{1}{J} + \frac{\chi}{J^2}\Bigr) + \frac{G}{J^{1/3}}\Omega =\mu_{o}\,, 
\end{equation}
where $R$ ([J/(K\,mol]) is the universal gas constant,  $T$ ([K]) is the temperature, $\chi$ is the Flory parameter, $G$ ([J/m$^3$]) is the shear modulus of the dry gel, and $\Omega$ ($[\Omega]=$m$^3$/mol) is the molar volume of the liquid. The change in volume is  equal to the liquid uptake:  $J=1+\Omega\,c$, with $c$ the liquid concentration  per unit of dry volume ([mol/m$^3$]. From now on we denote with  $J_o$ the change in volume from the dry to the swollen state $\Sc_o$. For $G = 10^4$ Pa, $\chi = 0.4$, $\Omega = 1.8\cdot 10^{-5}$ m$^3$/mol and $\mu_o=0$ J/mol, $J_o=77.6$, \textit{i.e.} it corresponds to an increase of the original dry volume $V_d$ of $77.6$ times, due to liquid uptake (see cartoon in figure \ref{fig:2}).\\
A further change of the environmental conditions, both mechanical and chemical, drives the gel to a new equilibrium state $\Sc$ and determines a further change in volume. For example, liquid is expelled if the chemical potential of the bath changes from $\mu_o$ to $\mu_{ext}<\mu_o$ and the gel shrinking is measured by the change in volume (from the dry state) known from equation \eqref{steady1} with $\mu_o$ changed in $\mu_{ext}$. 
For $\mu_{ext}=-0.123$ J/mol,  we get $J=38.8$, corresponding to halving the initial $\Sc_o$ volume, 
and thus releasing a liquid volume equal to $38.8$ times the dry volume $V_d$. 
It is worth noting that: (i) for $\mu_o\simeq 0$ ($1/\mu_o\simeq 0$), any small variations in the chemical potential determine high (small) changes in volume; (ii) these latter are driven by changes in the chemical potential of the bath which enter the model via chemical boundary conditions. Alternatively, the liquid content in passive gels can be changed by loading gels on their boundary\cite{JAP:2016}. 

\begin{figure}[t]
\includegraphics[width=0.45\textwidth]{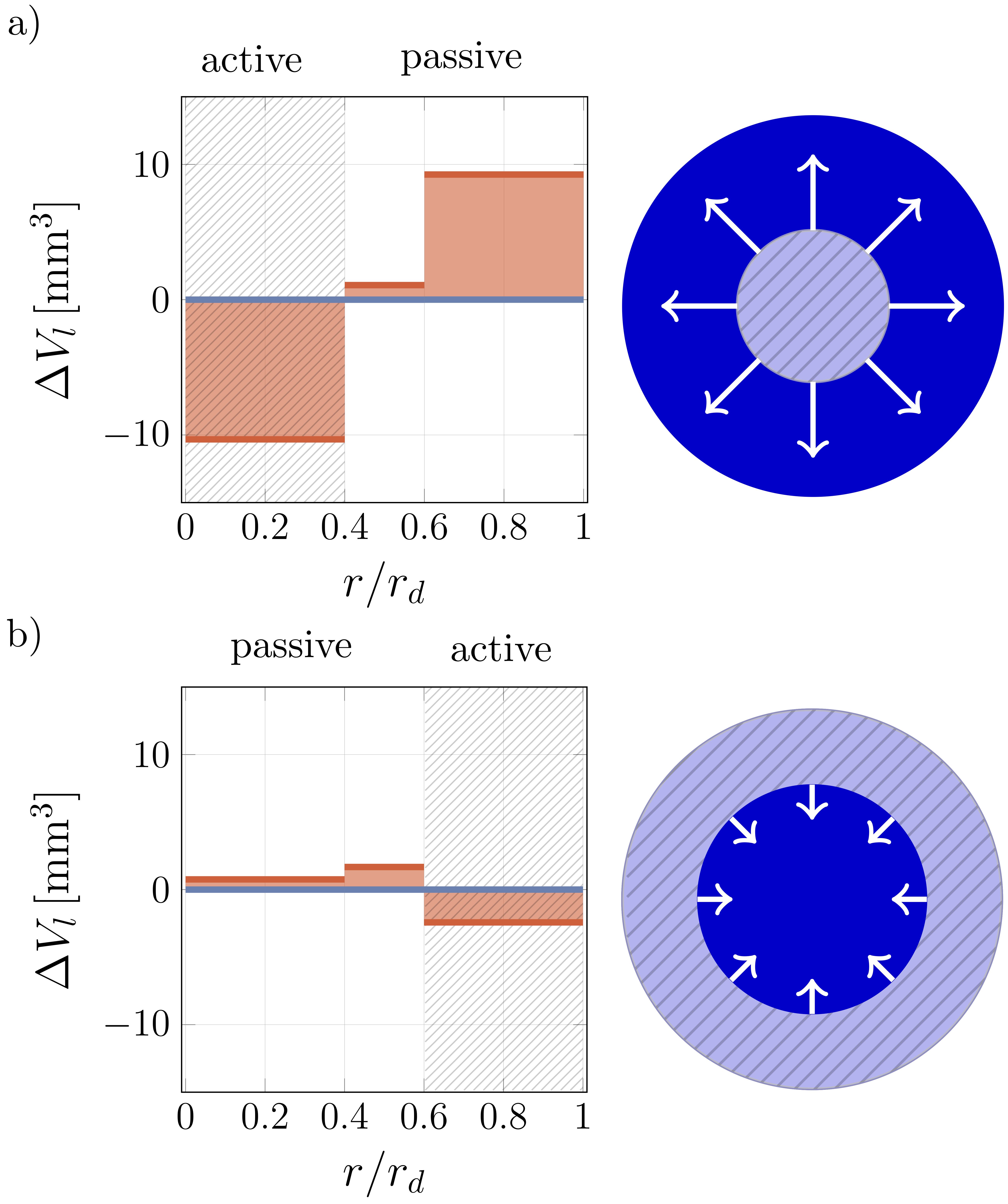}
\caption{\label{fig:4} Contractile activation of the impermeable spherical gel. 
Left panels show the volumes of liquid which are displaced due to activation: initial liquid distribution (blue line), final average 
liquid content (red line); the orange areas show the volume $\Delta V_l$ of liquid that has migrated.
Right panels show the active regions (light blue) and the passive regions (dark blue) in the sphere. Arrows denote liquid flux from the active contracted region to the passive one.
}
\end{figure}

\begin{figure}[h]
\includegraphics[width=0.7\textwidth]{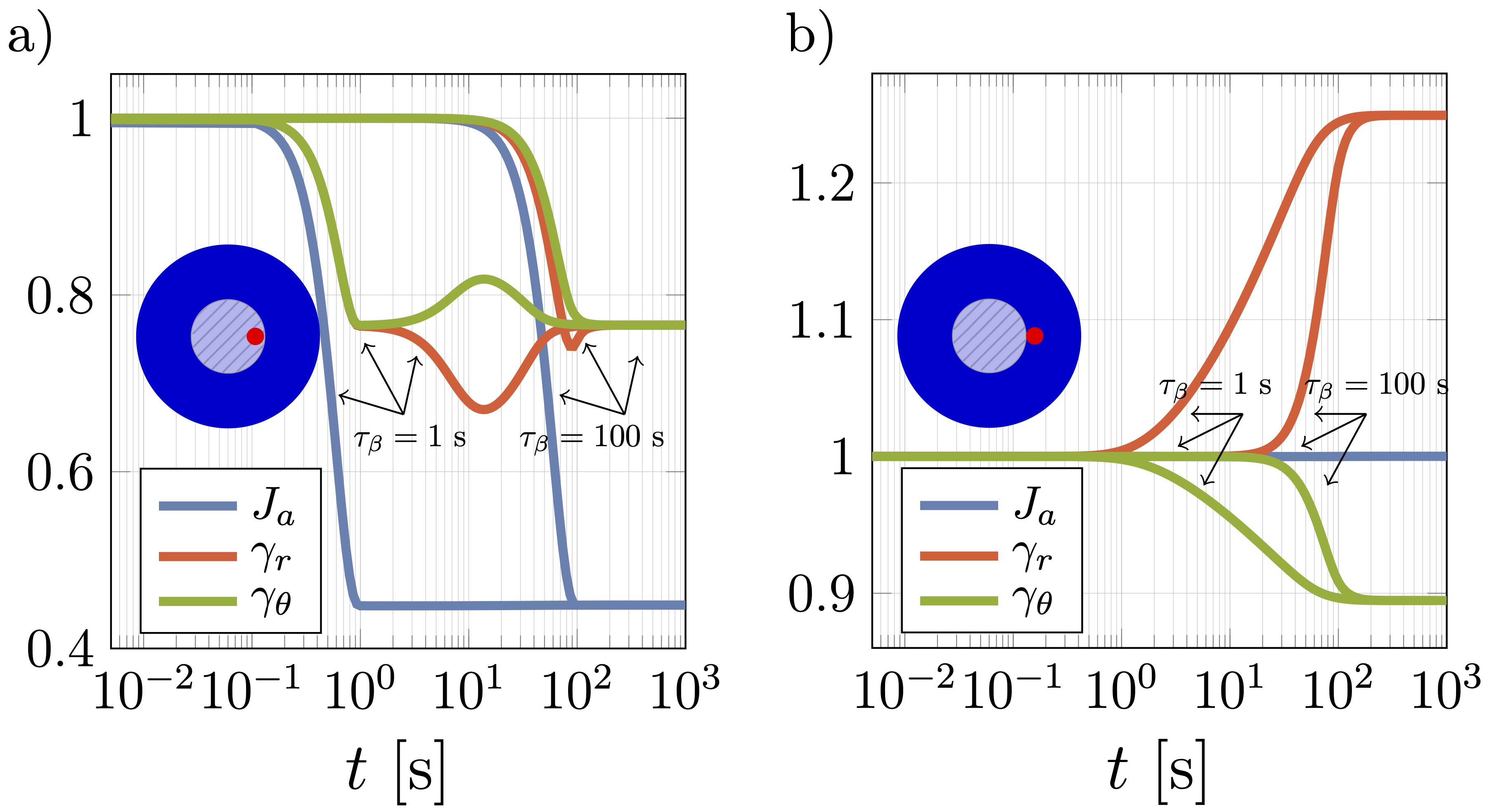}
\caption{\label{fig:6} Radial $\gamma_r$ and hoop $\gamma_\theta$ remodeling components (orange and green lines, respectively) and $J_a$ (blue lines) over time in two selected points (red points) inside (panel a) and outside (b) the active core region due to an increase in the remodeling external source  which takes the value $\Delta^c$ in a time $\tau_\beta=1$ s and $\tau_\beta=100$ s.}
\end{figure}

In the following, we shall show how an active gel can achieve a same de-swelling, remaining at constant the chemical potential, but
 shortening the mean free length of the polymer meshwork. \\
As a difference, the state of an active gel can be driven by bulk  through the action of inner (molecular) motors and the amount of energy required to attain the thermodynamical equilibrium state $\Sc$ under the same bath ($\mu_{ext}=\mu_o$) can be quantified through our model. Besides, it can describe the dynamics of liquid redistribution within the swollen state $\Sc_o$ under not uniform bulk actions in presence of impermeable boundaries (see figures \ref{fig:4}-\ref{fig:6}).\\
The chemo-mechanical state of an active gel sphere is described by the radial
displacement $u=u(r,t)$ from the dry state ([m]) and the liquid concentration $c=c(r,t)$ per unit of dry volume ([mol/m$^3$]), as for passive gels, plus 
the remodeling variables $\lmg_r=\lmg_r(r,t)$ and $\lmg_\theta=\lmg_\theta(r,t)$ which  describe the radial and hoop macroscopic changes in length of the sphere due to network remodeling.
The chemo-mechanical state of the active gel sphere is ruled by the balance of forces and liquid mass
\begin{equation}\label{bal}
\sigma_r^\prime + \frac{2}{r}(\sigma_r-\sigma_\theta) = 0\quad\textrm{and}\quad
\dot c = -(h^\prime +\frac{2}{r}\,h)\,,
\end{equation}
where $\sigma_r$ and  $\sigma_\theta$ are the radial and hoop components of the dry-reference stress and $h$ is the radial liquid flux. Moreover, the active gel sphere has two more balance laws\footnote{See \cite{Curatolo:2017_M,Curatolo:2019} for further details.}
\begin{equation}\label{remo}
m\,\frac{\dot \lmg_r}{\lmg_r}=\beta_r-E_r\quad\textrm{and}\quad
m\,\frac{\dot \lmg_\theta}{\lmg_\theta}=\beta_\theta-E_\theta\,,
\end{equation}
showing that the evolution of $\lmg_r$ and $\lmg_\theta$ is driven by the difference between the radial and hoop components $\beta_r$ and $\beta_\theta$ ([J/m$^3$])
of the remodeling bulk source and the corresponding Eshelby components $E_r$ and $E_\theta$ ([J/m$^3$]) \cite{DiCarlo:2002,Gurtin:2000,Curatolo:2017_M}, and depends on the resistance to remodeling $m$ ($[m]=$J s/m$^3$). 
The evolution of $\lmg_r$ and $\lmg_\theta$ depends on the remodeling sources and on the resistances to mobility $m$, which we assumed equal in the radial and hoop direction. 
A characteristic remodeling time $\tau_r$ is naturally identified by the ratio between $m$ and the intensity of the remodeling sources: 
$\tau_r=\textrm{min}(m/\vert\beta_r\vert,m/\vert\beta_\theta\vert)$.\\
For active gels, the change $J$ of volume due to swelling is equal to the liquid uptake or release up to the change 
$J_a=\lmg_r\lmg_\theta^2$ of the gel volume  due to remodeling
\begin{equation}\label{vc}
J = J_a + \Omega\,c\,.
\end{equation}
Equation \eqref{vc} dictates that liquid uptake locally determines the change of  volume  from the remodeled state to the actual state (see figure \ref{fig:1}). It also holds $J=\lambda_r\lambda_\theta^2$ with  the radial and hoop deformations  $\lambda_r=1+u'$ and  $\lambda_\theta=1+u/r$,
measured from the dry state. 

Thermodynamically consistent constitutive equations for $\sigma_r$ and $\sigma_\theta$,  $h$,  $E_r$ and $E_\theta$, and the chemical potential $\mu$ of the liquid within the gel 
are derived from a revisited Flory Rehner free-energy
\begin{equation}
\psi=J_a(\varphi_e+\varphi_m)-p\,(J-(J_a + \Omega c))
\end{equation}
per unit dry volume, additively split into the elastic component $\varphi_e$ and the mixing component $\varphi_m$ per unit remodeled volume, and accounting for the volumetric constraint \eqref{vc} maintained by the reaction $p$.
\footnote{
The Jacobians $J_a$ and $J$ have a key role in changing volume-elements. Precisely,
let $dV_d$, $dV_a$, and $dv$ be the dry-reference, the remodeled, and the actual volume-elements, respectively. Then, they are related by the Jacobians as
\begin{equation}\label{vols}
dV_a=J_a\,dV_d\quad\textrm{and}\quad 
dv  =\Jd\,dV_d = \frac{J}{J_a}\,dV_a.
\end{equation}
In these terms,  equation \eqref{vc} can be rewritten as 
\begin{equation}\label{sol}
dv=dV_a+dV_{\rm{sol}}\,,
\end{equation}
with $dV_{\rm{sol}}=\Omega\,c\,dV_d$ denoting the liquid volume-element. So, using these volume relationships, free energies per unit remodeled volume as $\varphi_e$ and $\varphi_m$ are transformed into free energies per unit dry volume as $\psi\,dV_d= (\varphi_e +\varphi_m)\,dV_a= J_a(\varphi_e +\varphi_m)\,dV_d$.} 
The elastic component $\varphi_e$ has a neo-Hookean form and depends on the effective radial and hoop strains $\lambda_r/\lmg_r$ and $\lambda_\theta/\lmg_\theta$, respectively\footnote{An effective strain measures a change in length from the remodeled state.}:
\begin{equation}
\varphi_e= \frac{G}{2}J_a\,\left(
\left(\lambda_r/\lmg_r\right)^2
      +2\,\left(\lambda_\theta/\lmg_\theta\right)^2 -3\right)\,.
\end{equation}
The mixing component $\varphi_m$ depends on the  \emph{polymer fraction} $\phi$, a key chemical variable, which is given by
\begin{equation}\label{VolFra}
\phi=\frac{J_a}{J}=\frac{J_a}{J_a+\Omega\,c}=\frac{1}{J_e}\,,
\quad\textrm{where } J_e=\frac{J}{J_a}\,,
\end{equation}
and takes the following form
\begin{equation}
\varphi_m=\frac{R\,T}{\Omega}\,\left( (J_e-1)\,\log\left(1-1/J_e\right) 
                                  + \chi\,\left(1-1/J_e\right) \right)\,,
\end{equation}
which is formally analogous to the standard Flory-Rehner mixing energy, proviso $J_e$ is replaced by $1/\phi$. 
However, from the point of view of macroscopic mechanics, it is quite different as here polymer fraction is not related to the visible change of volume $J$ (see equations \eqref{vc} and\eqref{VolFra}).
Standard constitutive procedures yields:
\begin{equation}\label{stresses}
\sigma_r=G\,\lambda_r\,\frac{\lmg_\theta^2}{\lmg_r} - p\,\lambda_\theta^2\quad\textrm{and}\quad
\sigma_\theta=G\,\lambda_\theta\,\lmg_r - p\,\lambda_r\,\lambda_\theta\,;
\end{equation}
\begin{equation}\label{mu}
\mu=R\,T\Bigl(\textrm{log}\frac{J_{e}-1}{J_{e}} + \frac{1}{J_{e}} + \frac{\chi}{J_{e}^2}\Bigr) 
            + p\,\Omega\,,
\end{equation}
\begin{equation}\label{h}
h=-\frac{D \, c}{R\,T\,\lambda_r^2}\,\mu'\,,
\end{equation}
with $D$ the diffusivity parameter ($[D]=$m$^2$/s).
The dynamics described by the equation \eqref{bal}$_2$ introduces the characteristic diffusion time $\tau_d=l^2/D\varepsilon$ into the model, where $l$ is  a characteristic diffusion length and 
$\varepsilon=G\Omega/RT$ measures the ratio between the elastic and the mixing energy. 
Finally, the radial and hoop Eshelby components are:
\begin{equation}\label{Ert}
E_r=\psi-\lambda_r\, \sigma_r+-c\,\mu\,,\,
E_\theta=\psi-\lambda_\theta\, \sigma_\theta-c\,\mu \,.
\end{equation}
%
%
For a given $\mu_o$, the stress--free solution $J_o$ of \eqref{steady1}
is also a steady solution if $J_a=1$ and \eqref{remo} yields $\dot\gamma_r=\dot\gamma_\theta=0$;
this second requirement implies that the resistance to mobility $m$ be infinite, or that $\beta_r=E_r$ and $\beta_\theta=E_\theta$.
It means that for active gels,  a remodeling source is required to  maintain the
thermodynamic equilibrium at $\Sc_o$ (unless $m\to\infty$)\cite{Curatolo:2019,Bacca:2019} .
Thus, at $\Sc_o$, we have $\sigma_r=\sigma_\theta=0$ and $\mu=\mu_o=0$ to fulfill the stress-free condition,
and $E_r=E_\theta=\beta_r=\beta_\theta=\beta_o=\psi_o$ for the steadiness, with $\psi_o$ the amount of energy per unit dry volume 
required to attain the thermodynamical equilibrium. In the present case, we get $\beta_o=-8\cdot 10^7$ J/m$^3$.
%
%

\noindent
The remodeling source is a further control of the state of the system: fixed $\mu_{ext}=\mu_o$,
it is possible to induce a liquid release or uptake  by changing the value of $\beta$. Following our example,
to $\beta_1=\beta_o+\Delta\beta$, with $\Delta\beta=4\cdot 10^7$ J/m$^3$, there correspond a different stress-free and steady state
$\Sc_1$, having $J_a=0.5$; it follows a halving of the initial volume, that is, $J=J_o J_a=38.8$ times the dry volume $V_d$. 
It is worth confronting this result with the aforementioned one for passive gel, where 
the halving of the volume has been induced by lowering  $\mu_{\rm{ext}}$

By contrast, for a $\Delta\beta<0$, we get a volume increase, as shown in figure \ref{fig:2}.
It is worth noting that the final state only
depends on $\Delta\beta$, but the time evolution of the state variables is related to the time $\tau_\beta$ 
required to switch from $\beta_o$ to $\beta_1$.

\begin{figure}[h]
\includegraphics[width=0.7\textwidth]{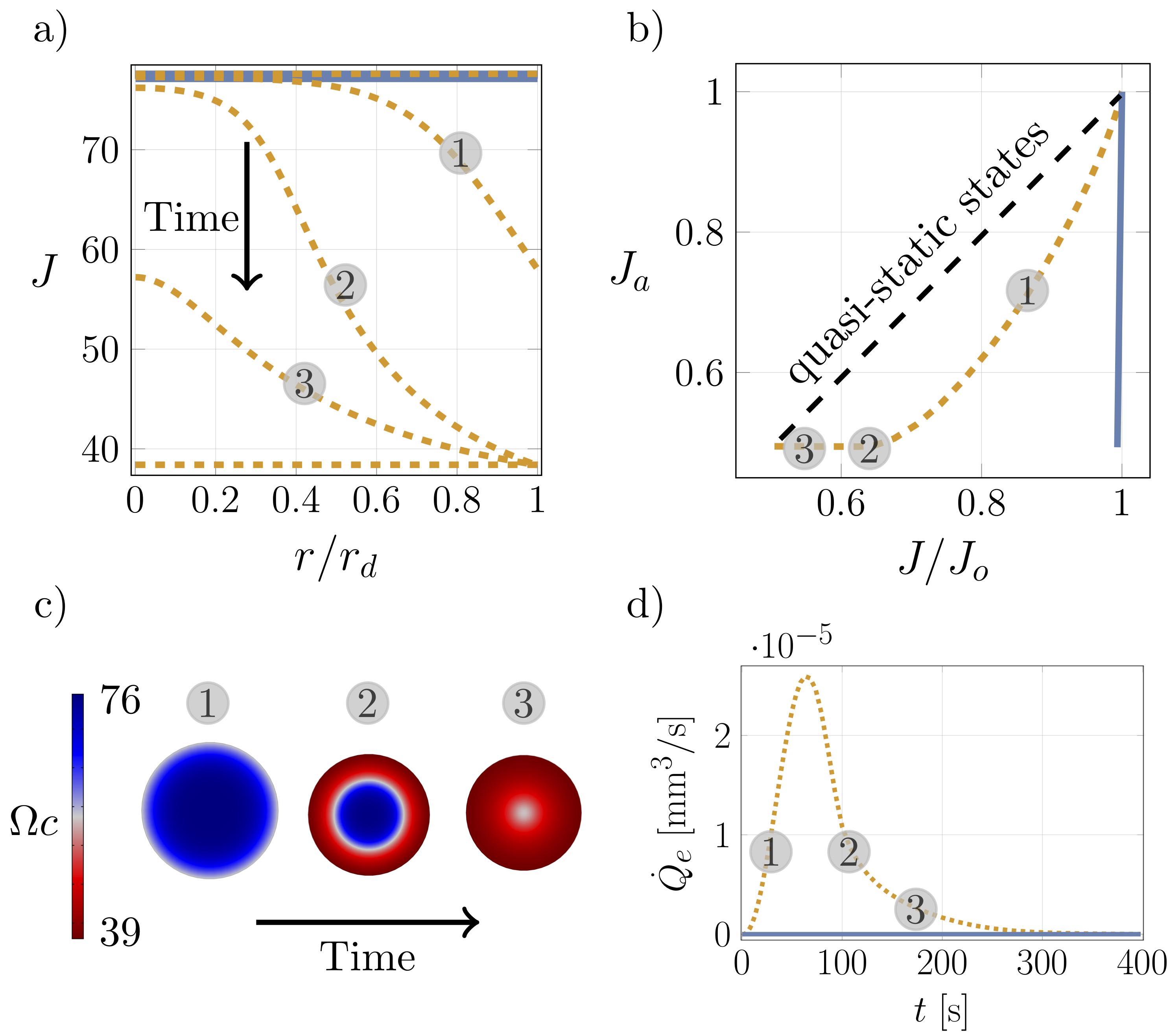}
\caption{\label{fig:3} Contraction of a permeable spherical gel. 
(a) Swollen volume ratio  $J$ versus dimensionless radius $r/r_d$  (dashed orange line) at three instants during time evolution; deswelling starts at the boundary and becomes uniform at steady state; $t_1=50$ s, $t_2=100$ s, $t_3=150$ s.
(b) Trajectory of the point $(\bar J(t)/J_o,\bar J_a(t))$ (dashed orange line) in the  $(J/J_o, J_a)$ plane;  the dashed black line represents the stress-free states. 
(c) Liquid content at the three instants of panel (b). 
(d) Liquid flux:  after the initial peak it decreases until the steady value is attained, when liquid flux is null.
The case of the impermeable sphere is also shown in blue solid lines.} 
\end{figure}

The dynamics from $\Sc_o$ to $\Sc_1$ is ruled by the equations \eqref{bal}-\eqref{remo}, and 
the results corresponding to $\tau_\beta=100$ s, $m=10^5$ Pa/s, $D=10^{-3}$ m$^2$/s, and a sphere of dry radius $r_d=1$ mm,
are described in figure \ref{fig:3}.  Panel (a) shows the local volume change $J$ versus the dimensionless  radius $r/r_d$ (orange dashed lines), at three times $t_i$; the corresponding liquid distribution  is shown in the bottom panel (c). 
At  $t_2=100$ s,  with $t_2 > \tau_d \simeq 10^1$ s $>>\tau_r=10^{-3}$ s, the liquid has already been completely expelled from the 
periphery of the sphere, where $J=38.8$, whereas in the core it still takes the initial value $J_o$.
%
%
%
%
%

The trajectory of the mean values $\bar J$, $\bar J_a$
of the volume ratios can be illustrated in the contraction-swelling diagram of axes $J/J_o$ and $ J_a$ in panel (b). As  stresses are zero and $\mu=\mu_o$ at both $\Sc_o$ and $\Sc_1$, $J_e$ takes the same value at the two states (see also figure \ref{fig:2}). Hence, as  $J_e=J_o$ at $\Sc_o$ where $J_a=1$ and $J=J_o$, the point $(1,1)$ in the diagram marks $\Sc_o$; at any other equilibrium state,  $J_e=J_o=J/J_a$. It identifies the  bisectrix of the diagram (dashed black line) as the locus of the stress-free states. So, the trajectory of $\bar J$ and $\bar J_a$  starts and ends on the line and is far from it along the transient (orange dashed lines). Finally, the liquid flux $\dot Q_e=\Omega h$ ($[\dot Q_e]=$mm$^3$/s) 
(panel d) shows a pattern similar to the one already evidenced when contraction of swollen active soft matter is involved \cite{Bernheim:2018_n,Curatolo:2019,Bacca:2019}, with a peak at the first times and slowly decreases to zero.\\
When the boundary of the swollen sphere is made impermeable, liquid is exchanged with the environment
and $J$ changes only because of the contraction $J_a$, see blue solid lines in top panels of figure \ref{fig:3}. 
The trajectory of the mean values $\bar J$, $\bar J_a$ is almost vertical (panel b),  
and ends at a highly stressed state.  To get liquid migration within the swollen state $\Sc_o$,  
a not uniform bulk active source is required.\\

In particular, we study liquid migration as consequence of a stepwise constant remodeling source which segregates the sphere into 
an active and a passive region where $\beta>\beta_o$ and  $\beta=\beta_o$, respectively. 
We consider both a core activation and a peripheral activation:
$\beta=\beta_o+\Delta^c$  in the core region $0\le r/r_d<r_c=0.4$
and $\beta=\beta_o$ otherwise,
$\beta=\beta_o+\Delta^p$ in the peripheral region $0.6=r_p \le r/r_d \le 1$
and $\beta=\beta_o$ otherwise, respectively.
%
%
%
%
%
The intensities $\Delta^c$ and $\Delta^p$ are chosen in such a way to have the same amount of energy $\Delta^c\,V_c=\Delta^p\,V_p = 0.012$ J, with $V_c=4/3\pi r_c^3$ and $V_p=4/3\pi(r_d^3-r_p^3)$. Our results, summed up in figures \ref{fig:4}-\ref{fig:6}, allow to discuss the interaction between liquid migration and stress generation in the dynamics of active gels.\\
The starting point of the analysis is $\Sc_o$ and we assume that the dry sphere swelled and only later its boundary was made impermeable so that the uptaken liquid is trapped into $\Sc_o$ forever. At $\Sc_o$, the liquid volume $V_{lo}$ is $J_o\,V_d=3.25\cdot 10^{-7}$ m$^3$ and the steady state is maintained by the bulk source $\beta_o$.  Any change $\Delta V_l$ in the liquid volume from $V_{lo}$ is due to the successive increases in the remodeling source. 

Figure \ref{fig:4} shows the initial, uniform liquid distribution in the $r/r_d-\Delta V_l$ plane as a blue solid line ($\Delta V_l=0$). Red lines denote the average 
liquid content at the final state, in the core and in the intermediate and peripheral regions; thus, the orange areas represent the volume of liquid
which has migrated. When the core is activated (top), a large volume of liquid migrates from the core, where $\Delta V_l<0$, to the 
periphery where $\Delta V_l>0$, with an almost negligible change of the liquid content in the intermediate region. 
On the contrary, in the case of peripheral activation of equal global intensity and hence smaller density, a small migration towards the core is realized, which mainly involves both the intermediate region and the core. The cartoon on the right depict active regions in light blue and passive regions in dark blue; arrows denote liquid flux from the active to the passive regions.\\
%
%

\begin{figure}[h]
\includegraphics[width=0.7\textwidth]{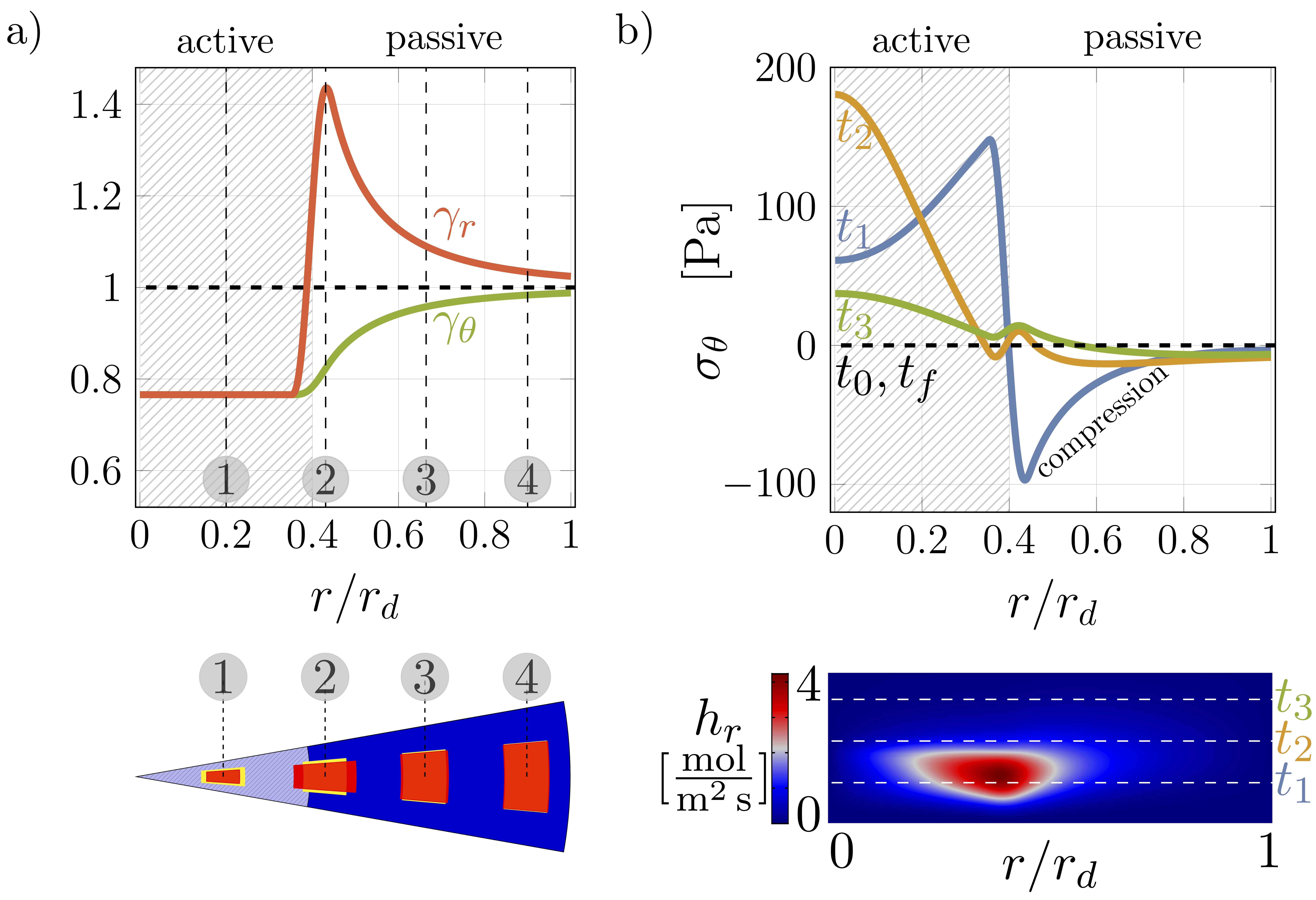}
\caption{\label{fig:5} Remodeling components at steady final state and stress dynamics for the active core case corresonding to $\tau_\beta=100$s. a) Top: remodeling components $\gamma_r$,$\gamma_\theta$ at steady final state have same value in the active region, fully isotropically contracted, while have different values in the passive region. Bottom: cartoon with four spherical volume elements at initial and final state, respectively yellow and red colored. The shape change of the volume element due to remodeling processes are necessary to realize a stress-free final state. b) Top: High stresses are reached firstly close to the active/passive interface (time $t_1=50$s), then higher stresses move towards the center (times $t_2=100$s and $t_3=150$s. Black dashed line represents the initial and final state which are stress-free. Bottom: flux over time is high close to the interface while is low at the center nevertheless high stresses are reached at same instants.}
\end{figure}

\noindent
The model allows to look at the pattern over time of the radial and hoop remodeling variables $\gamma_r$ and $\gamma_\theta$ in two points at the interface between active and passive regions: one inside the core (figure \ref{fig:6}, left panel) and one outside the core  (figure \ref{fig:5}, right panel). They describe the macroscopic changes in length of the sphere due to network remodeling and are not visible unless they are realized and take the values $\lambda_r$ and $\lambda_\theta$ of the radial and hoop deformations. In this case, it would hold $J_a=J$, no changes in the liquid concentration would occur (see equation \eqref{vc}) and stresses would be zero. The liquid content of the sphere makes this situation unrealizable: material remodeling determines liquid migration which is coupled to the stress state in the sphere.\\
 The patterns $\gamma_r(t)$ and $\gamma_\theta(t)$ are driven by the equations \eqref{remo} and \eqref{Ert} and, due to the isotropic mobility and activation source and granted for the representation form of the Eshelby components, may differ one from each other only for the  $\lambda_r\sigma_r$ and $\lambda_\theta\sigma_\theta$ components  in the equations  \eqref{Ert}.
Inside the active core, due to the difference between remodeling and diffusion characteristic times 
($\tau_r\simeq 10^{-3}\,\rm{s} <<\tau_d\simeq 10$ s) and being $\tau_r<<\tau_\beta$ with $\tau_\beta=1$ s, or 100 s,  before diffusion starts $\gamma_r$ and $\gamma_\theta$   are strongly driven by the increase $\Delta^c(t)$ in the activation source, and take their steady values when $\Delta^c=\Delta^c(t_\beta)$. Moreover, when diffusion starts liquid starts being released from the core so driving a negative radial stress and a positive hoop stress which determine the characteristic patterns observed in figure \ref{fig:6}, which disappear when $\tau_d$ is over. These behaviours are more evident for fast  ($\tau_\beta=1$s) than for slow ($\tau_\beta=10^2$ s) activation sources.\\ 
Actually, figure \ref{fig:6} also records $J_a=\gamma_r\gamma_\theta^2$ over time which delivers changes in volume due to activation.
It confirms that the remodeling-induced change in the gel volume in the active core is very fast and we can observe an almost direct relationship between $\Delta^c(t)$ and $J_a(t)$ which takes its steady value  $J_a\approx0.5$ in the time $\tau_\beta$. This value is unchanged by diffusion, due to commented above changes in $\gamma_r$ and $\gamma_\theta$ induced by diffusion.\\
On the other hand, in the passive region before diffusion starts remodeling is absent and both the components $\gamma_r$, $\gamma_\theta$ and $J_a$ keep the unit value. When diffusion starts, due to the activation of the core, liquid starts migrating from the core to the pheripery, also remodeling starts and $\gamma_r$ and $\gamma_\theta$ take different patterns: a radial expansion up to $\gamma_r\approx 1.25$ and a hoop contraction up to $\gamma_\theta\approx 0.9$, corresponding to a positive radial stress and a negative radial stress, as it is expected in a region whereas liquid has been uptaking.\\
The values taken by $\gamma_r$ and $\gamma_\theta$ at the steady stress-free state are shown in  the left panel of figure \ref{fig:5} (top),  together with the remodeling of four volume elements along the radius of the sphere (bottom) for $\tau_\beta=100$s . In the core region, volume elements are isotropically contracted (point $1$); at the interface, the difference between $\gamma_r$ and $\gamma_\theta$ makes the contraction of the volume element highly anisotropic (point $2$); after the peak, the anisotropy of the contraction reduces moving towards the boundary (point $3$) up to a minimum value at point $4$. 
At the interface, as also shown by figure \ref{fig:6}, radial remodeling quickly changes from contraction (in the core) to expansion (in the periphery). As it corresponds to a stress-free state, it can be easily shown from equations \eqref{stresses} that  $\gamma_r$ and $\gamma_\theta$ also defines the visible radial and hoop deformations $\lambda_r$ and $\lambda_\theta$ up to a common multiplier. Hence, as expected, figure \ref{fig:5} reveals that at the interface sphere deformation is anisotropic and  corresponds to a radial contraction/expansion at the active/passive region whereas it moves towards isotropy at the boundary.
%
%
The hoop stress $\sigma_\theta$ evolves in time (see figure \ref{fig:5}, right top panel) to accomodate liquid migration  
which after a time $t_2>\tau_d$ has decreased significantly, as shown in figure \ref{fig:5}, see  panel (d).\\ 
In summary, bulk contraction can be modulated to get liquid migration and material and geometrical parameters can be tuned to realize fast contraction and delayed liquid migration. 

The model is able to describe different phenomena where water motion and growth or contraction of a solid are involved. The present study delivers a deep insight into processes that can take place in natural elements, such as cells, but also in synthetic active gel system. The results for an impermeable gel sphere demonstrate that the migration of liquid, which is essential in biological organisms and tissues, is strongly coupled with remodeling actions. The latter, are necessary not only to relax stresses but also to induce a change of the body shape which is often observed in natural systems.  Moreover, we also show as differential growth/contraction patterns are realized both in regions where gel is activated but also where liquid migrates. This observation opens new possibilities to predict the biological behavior in cells and tissues where high stresses can lead to disruptive processes at micro and macro scale.

This work is supported by MIUR (Italian Minister for Education, Research, and University) 
through \emph{PRIN 2017, Mathematics of active materials: From mechanobiology to smart devices},
project n. 2017KL4EF3.
%
%
\bibliography{2019_gel}

\begin{thebibliography}{22}%
\makeatletter
\providecommand \@ifxundefined [1]{%
 \@ifx{#1\undefined}
}%
\providecommand \@ifnum [1]{%
 \ifnum #1\expandafter \@firstoftwo
 \else \expandafter \@secondoftwo
 \fi
}%
\providecommand \@ifx [1]{%
 \ifx #1\expandafter \@firstoftwo
 \else \expandafter \@secondoftwo
 \fi
}%
\providecommand \natexlab [1]{#1}%
\providecommand \enquote  [1]{``#1''}%
\providecommand \bibnamefont  [1]{#1}%
\providecommand \bibfnamefont [1]{#1}%
\providecommand \citenamefont [1]{#1}%
\providecommand \href@noop [0]{\@secondoftwo}%
\providecommand \href [0]{\begingroup \@sanitize@url \@href}%
\providecommand \@href[1]{\@@startlink{#1}\@@href}%
\providecommand \@@href[1]{\endgroup#1\@@endlink}%
\providecommand \@sanitize@url [0]{\catcode `\\12\catcode `\$12\catcode
  `\&12\catcode `\#12\catcode `\^12\catcode `\_12\catcode `\%12\relax}%
\providecommand \@@startlink[1]{}%
\providecommand \@@endlink[0]{}%
\providecommand \url  [0]{\begingroup\@sanitize@url \@url }%
\providecommand \@url [1]{\endgroup\@href {#1}{\urlprefix }}%
\providecommand \urlprefix  [0]{URL }%
\providecommand \Eprint [0]{\href }%
\providecommand \doibase [0]{http://dx.doi.org/}%
\providecommand \selectlanguage [0]{\@gobble}%
\providecommand \bibinfo  [0]{\@secondoftwo}%
\providecommand \bibfield  [0]{\@secondoftwo}%
\providecommand \translation [1]{[#1]}%
\providecommand \BibitemOpen [0]{}%
\providecommand \bibitemStop [0]{}%
\providecommand \bibitemNoStop [0]{.\EOS\space}%
\providecommand \EOS [0]{\spacefactor3000\relax}%
\providecommand \BibitemShut  [1]{\csname bibitem#1\endcsname}%
\let\auto@bib@innerbib\@empty
\bibitem [{\citenamefont {Moeendarbary}\ \emph {et~al.}(2013)\citenamefont
  {Moeendarbary}, \citenamefont {Valon}, \citenamefont {Fritzsche},
  \citenamefont {Harris}, \citenamefont {Moulding}, \citenamefont {Thrasher},
  \citenamefont {Stride}, \citenamefont {Mahadevan},\ and\ \citenamefont
  {Charras}}]{Moeenda:2013}%
  \BibitemOpen
  \bibfield  {author} {\bibinfo {author} {\bibfnamefont {E.}~\bibnamefont
  {Moeendarbary}}, \bibinfo {author} {\bibfnamefont {L.}~\bibnamefont {Valon}},
  \bibinfo {author} {\bibfnamefont {M.}~\bibnamefont {Fritzsche}}, \bibinfo
  {author} {\bibfnamefont {A.~R.}\ \bibnamefont {Harris}}, \bibinfo {author}
  {\bibfnamefont {D.~A.}\ \bibnamefont {Moulding}}, \bibinfo {author}
  {\bibfnamefont {A.~J.}\ \bibnamefont {Thrasher}}, \bibinfo {author}
  {\bibfnamefont {E.}~\bibnamefont {Stride}}, \bibinfo {author} {\bibfnamefont
  {L.}~\bibnamefont {Mahadevan}}, \ and\ \bibinfo {author} {\bibfnamefont
  {G.~T.}\ \bibnamefont {Charras}},\ }\bibfield  {title} {\enquote {\bibinfo
  {title} {The cytoplasm of living cells behaves as a poroelastic material},}\
  }\href {\doibase 10.1038/nmat3517} {\bibfield  {journal} {\bibinfo  {journal}
  {Nature Materials}\ }\textbf {\bibinfo {volume} {12}},\ \bibinfo {pages}
  {253--261} (\bibinfo {year} {2013})}\BibitemShut {NoStop}%
\bibitem [{\citenamefont {Prost}, \citenamefont {Julicher},\ and\ \citenamefont
  {Joanny}(2015)}]{Prost:2015}%
  \BibitemOpen
  \bibfield  {author} {\bibinfo {author} {\bibfnamefont {J.}~\bibnamefont
  {Prost}}, \bibinfo {author} {\bibfnamefont {F.}~\bibnamefont {Julicher}}, \
  and\ \bibinfo {author} {\bibfnamefont {J.-F.}\ \bibnamefont {Joanny}},\
  }\bibfield  {title} {\enquote {\bibinfo {title} {Active gel physics},}\
  }\href {\doibase http://dx.doi.org/10.1016/j.ijnonlinmec.2014.05.007}
  {\bibfield  {journal} {\bibinfo  {journal} {Nature Physics}\ }\textbf
  {\bibinfo {volume} {11}},\ \bibinfo {pages} {111 -- 117} (\bibinfo {year}
  {2015})},\ \bibinfo {note} {mechanics of Rubber - in Memory of Alan
  Gent}\BibitemShut {NoStop}%
\bibitem [{\citenamefont {Armstrong}\ \emph {et~al.}(2016)\citenamefont
  {Armstrong}, \citenamefont {Buganza~Tepole}, \citenamefont {Kuhl},
  \citenamefont {Simon},\ and\ \citenamefont {Vande~Geest}}]{Armstrong:2016}%
  \BibitemOpen
  \bibfield  {author} {\bibinfo {author} {\bibfnamefont {M.~H.}\ \bibnamefont
  {Armstrong}}, \bibinfo {author} {\bibfnamefont {A.}~\bibnamefont
  {Buganza~Tepole}}, \bibinfo {author} {\bibfnamefont {E.}~\bibnamefont
  {Kuhl}}, \bibinfo {author} {\bibfnamefont {B.~R.}\ \bibnamefont {Simon}}, \
  and\ \bibinfo {author} {\bibfnamefont {J.~P.}\ \bibnamefont {Vande~Geest}},\
  }\bibfield  {title} {\enquote {\bibinfo {title} {A finite element model for
  mixed porohyperelasticity with transport, swelling, and growth},}\ }\href
  {\doibase 10.1371/journal.pone.0152806} {\bibfield  {journal} {\bibinfo
  {journal} {PLOS ONE}\ }\textbf {\bibinfo {volume} {11}},\ \bibinfo {pages}
  {1--35} (\bibinfo {year} {2016})}\BibitemShut {NoStop}%
\bibitem [{\citenamefont {Curatolo}, \citenamefont {Gabriele},\ and\
  \citenamefont {Teresi}(2017)}]{Curatolo:2017_M}%
  \BibitemOpen
  \bibfield  {author} {\bibinfo {author} {\bibfnamefont {M.}~\bibnamefont
  {Curatolo}}, \bibinfo {author} {\bibfnamefont {S.}~\bibnamefont {Gabriele}},
  \ and\ \bibinfo {author} {\bibfnamefont {L.}~\bibnamefont {Teresi}},\
  }\bibfield  {title} {\enquote {\bibinfo {title} {Swelling and growth: a
  constitutive framework for active solids},}\ }\href {\doibase
  10.1007/s11012-017-0629-x} {\bibfield  {journal} {\bibinfo  {journal}
  {Meccanica}\ }\textbf {\bibinfo {volume} {52}},\ \bibinfo {pages}
  {3443--3456} (\bibinfo {year} {2017})}\BibitemShut {NoStop}%
\bibitem [{\citenamefont {Bacca}, \citenamefont {Saleh},\ and\ \citenamefont
  {McMeeking}(2019)}]{Bacca:2019}%
  \BibitemOpen
  \bibfield  {author} {\bibinfo {author} {\bibfnamefont {M.}~\bibnamefont
  {Bacca}}, \bibinfo {author} {\bibfnamefont {O.~A.}\ \bibnamefont {Saleh}}, \
  and\ \bibinfo {author} {\bibfnamefont {R.~M.}\ \bibnamefont {McMeeking}},\
  }\bibfield  {title} {\enquote {\bibinfo {title} {Contraction of polymer gels
  created by the activity of molecular motors},}\ }\href {\doibase
  10.1039/C8SM02598C} {\bibfield  {journal} {\bibinfo  {journal} {Soft Matter}\
  }\textbf {\bibinfo {volume} {15}},\ \bibinfo {pages} {4467--4475} (\bibinfo
  {year} {2019})}\BibitemShut {NoStop}%
\bibitem [{\citenamefont {Curatolo}, \citenamefont {Nardinocchi},\ and\
  \citenamefont {Teresi}(2020)}]{Curatolo:2019}%
  \BibitemOpen
  \bibfield  {author} {\bibinfo {author} {\bibfnamefont {M.}~\bibnamefont
  {Curatolo}}, \bibinfo {author} {\bibfnamefont {P.}~\bibnamefont
  {Nardinocchi}}, \ and\ \bibinfo {author} {\bibfnamefont {L.}~\bibnamefont
  {Teresi}},\ }\bibfield  {title} {\enquote {\bibinfo {title} {Dynamics of
  active swelling in contractile polymer gels},}\ }\href {\doibase
  https://doi.org/10.1016/j.jmps.2019.103807} {\bibfield  {journal} {\bibinfo
  {journal} {Journal of the Mechanics and Physics of Solids}\ }\textbf
  {\bibinfo {volume} {135}},\ \bibinfo {pages} {103807} (\bibinfo {year}
  {2020})}\BibitemShut {NoStop}%
\bibitem [{\citenamefont {Kroeger}, \citenamefont {Zerzour},\ and\
  \citenamefont {Geitmann}(2011)}]{Kroeger:2011}%
  \BibitemOpen
  \bibfield  {author} {\bibinfo {author} {\bibfnamefont {J.~H.}\ \bibnamefont
  {Kroeger}}, \bibinfo {author} {\bibfnamefont {R.}~\bibnamefont {Zerzour}}, \
  and\ \bibinfo {author} {\bibfnamefont {A.}~\bibnamefont {Geitmann}},\
  }\bibfield  {title} {\enquote {\bibinfo {title} {Regulator or driving force?
  the role of turgor pressure in oscillatory plant cell growth},}\ }\href
  {\doibase 10.1371/journal.pone.0018549} {\bibfield  {journal} {\bibinfo
  {journal} {PLOS ONE}\ }\textbf {\bibinfo {volume} {6}},\ \bibinfo {pages}
  {1--12} (\bibinfo {year} {2011})}\BibitemShut {NoStop}%
\bibitem [{\citenamefont {Beauzamy}, \citenamefont {Nakayama},\ and\
  \citenamefont {Boudaoud}(2014)}]{Beauzamy:2014}%
  \BibitemOpen
  \bibfield  {author} {\bibinfo {author} {\bibfnamefont {L.}~\bibnamefont
  {Beauzamy}}, \bibinfo {author} {\bibfnamefont {N.}~\bibnamefont {Nakayama}},
  \ and\ \bibinfo {author} {\bibfnamefont {A.}~\bibnamefont {Boudaoud}},\
  }\bibfield  {title} {\enquote {\bibinfo {title} {{Flowers under pressure: ins
  and outs of turgor regulation in development}},}\ }\href {\doibase
  10.1093/aob/mcu187} {\bibfield  {journal} {\bibinfo  {journal} {Annals of
  Botany}\ }\textbf {\bibinfo {volume} {114}},\ \bibinfo {pages} {1517--1533}
  (\bibinfo {year} {2014})},\ \Eprint
  {http://arxiv.org/abs/https://academic.oup.com/aob/article-pdf/114/7/1517/16993167/mcu187.pdf}
  {https://academic.oup.com/aob/article-pdf/114/7/1517/16993167/mcu187.pdf}
  \BibitemShut {NoStop}%
\bibitem [{\citenamefont {Sahaf}\ and\ \citenamefont
  {Sharon}(2016)}]{Sahaf:2016}%
  \BibitemOpen
  \bibfield  {author} {\bibinfo {author} {\bibfnamefont {M.}~\bibnamefont
  {Sahaf}}\ and\ \bibinfo {author} {\bibfnamefont {E.}~\bibnamefont {Sharon}},\
  }\bibfield  {title} {\enquote {\bibinfo {title} {{The rheology of a growing
  leaf: stress-induced changes in the mechanical properties of leaves}},}\
  }\href {\doibase 10.1093/jxb/erw316} {\bibfield  {journal} {\bibinfo
  {journal} {Journal of Experimental Botany}\ }\textbf {\bibinfo {volume}
  {67}},\ \bibinfo {pages} {5509--5515} (\bibinfo {year} {2016})},\ \Eprint
  {http://arxiv.org/abs/https://academic.oup.com/jxb/article-pdf/67/18/5509/18087327/2485673.pdf}
  {https://academic.oup.com/jxb/article-pdf/67/18/5509/18087327/2485673.pdf}
  \BibitemShut {NoStop}%
\bibitem [{\citenamefont {Zhang}\ and\ \citenamefont
  {Zhang}(2020)}]{Zhang:2020}%
  \BibitemOpen
  \bibfield  {author} {\bibinfo {author} {\bibfnamefont {D.}~\bibnamefont
  {Zhang}}\ and\ \bibinfo {author} {\bibfnamefont {B.}~\bibnamefont {Zhang}},\
  }\bibfield  {title} {\enquote {\bibinfo {title} {Pectin drives cell wall
  morphogenesis without turgor pressure},}\ }\href {\doibase
  https://doi.org/10.1016/j.tplants.2020.05.007} {\bibfield  {journal}
  {\bibinfo  {journal} {Trends in Plant Science}\ } (\bibinfo {year} {2020}),\
  https://doi.org/10.1016/j.tplants.2020.05.007}\BibitemShut {NoStop}%
\bibitem [{\citenamefont {Xu}, \citenamefont {Fu},\ and\ \citenamefont
  {Yang}(2020)}]{Xu:2020}%
  \BibitemOpen
  \bibfield  {author} {\bibinfo {author} {\bibfnamefont {F.}~\bibnamefont
  {Xu}}, \bibinfo {author} {\bibfnamefont {C.}~\bibnamefont {Fu}}, \ and\
  \bibinfo {author} {\bibfnamefont {Y.}~\bibnamefont {Yang}},\ }\bibfield
  {title} {\enquote {\bibinfo {title} {Water affects morphogenesis of growing
  aquatic plant leaves},}\ }\href {\doibase 10.1103/PhysRevLett.124.038003}
  {\bibfield  {journal} {\bibinfo  {journal} {Phys. Rev. Lett.}\ }\textbf
  {\bibinfo {volume} {124}},\ \bibinfo {pages} {038003} (\bibinfo {year}
  {2020})}\BibitemShut {NoStop}%
\bibitem [{\citenamefont {Fei}\ \emph {et~al.}(2020)\citenamefont {Fei},
  \citenamefont {Mao}, \citenamefont {Yan}, \citenamefont {Alert},
  \citenamefont {Stone}, \citenamefont {Bassler}, \citenamefont {Wingreen},\
  and\ \citenamefont {Ko{\v s}mrlj}}]{Fei:2020}%
  \BibitemOpen
  \bibfield  {author} {\bibinfo {author} {\bibfnamefont {C.}~\bibnamefont
  {Fei}}, \bibinfo {author} {\bibfnamefont {S.}~\bibnamefont {Mao}}, \bibinfo
  {author} {\bibfnamefont {J.}~\bibnamefont {Yan}}, \bibinfo {author}
  {\bibfnamefont {R.}~\bibnamefont {Alert}}, \bibinfo {author} {\bibfnamefont
  {H.~A.}\ \bibnamefont {Stone}}, \bibinfo {author} {\bibfnamefont {B.~L.}\
  \bibnamefont {Bassler}}, \bibinfo {author} {\bibfnamefont {N.~S.}\
  \bibnamefont {Wingreen}}, \ and\ \bibinfo {author} {\bibfnamefont
  {A.}~\bibnamefont {Ko{\v s}mrlj}},\ }\bibfield  {title} {\enquote {\bibinfo
  {title} {Nonuniform growth and surface friction determine bacterial biofilm
  morphology on soft substrates},}\ }\href {\doibase 10.1073/pnas.1919607117}
  {\bibfield  {journal} {\bibinfo  {journal} {Proceedings of the National
  Academy of Sciences}\ }\textbf {\bibinfo {volume} {117}},\ \bibinfo {pages}
  {7622--7632} (\bibinfo {year} {2020})},\ \Eprint
  {http://arxiv.org/abs/https://www.pnas.org/content/117/14/7622.full.pdf}
  {https://www.pnas.org/content/117/14/7622.full.pdf} \BibitemShut {NoStop}%
\bibitem [{\citenamefont {Flory}\ and\ \citenamefont
  {Rehner}(1943{\natexlab{a}})}]{FR1}%
  \BibitemOpen
  \bibfield  {author} {\bibinfo {author} {\bibfnamefont {P.~J.}\ \bibnamefont
  {Flory}}\ and\ \bibinfo {author} {\bibfnamefont {J.}~\bibnamefont {Rehner}},\
  }\bibfield  {title} {\enquote {\bibinfo {title} {Statistical mechanics of
  cross-linked polymer networks i. rubberlike elasticity},}\ }\href {\doibase
  10.1063/1.1723791} {\bibfield  {journal} {\bibinfo  {journal} {J Chem Phys}\
  }\textbf {\bibinfo {volume} {11}},\ \bibinfo {pages} {512--520} (\bibinfo
  {year} {1943}{\natexlab{a}})}\BibitemShut {NoStop}%
\bibitem [{\citenamefont {Flory}\ and\ \citenamefont
  {Rehner}(1943{\natexlab{b}})}]{FR2}%
  \BibitemOpen
  \bibfield  {author} {\bibinfo {author} {\bibfnamefont {P.~J.}\ \bibnamefont
  {Flory}}\ and\ \bibinfo {author} {\bibfnamefont {J.}~\bibnamefont {Rehner}},\
  }\bibfield  {title} {\enquote {\bibinfo {title} {Statistical mechanics of
  cross-linked polymer networks ii. swelling},}\ }\href {\doibase
  10.1063/1.1723792} {\bibfield  {journal} {\bibinfo  {journal} {J Chem Phys}\
  }\textbf {\bibinfo {volume} {11}},\ \bibinfo {pages} {521--526} (\bibinfo
  {year} {1943}{\natexlab{b}})}\BibitemShut {NoStop}%
\bibitem [{\citenamefont {Rodriguez}, \citenamefont {Hoger},\ and\
  \citenamefont {McCulloch}(1994)}]{Rodriguez:1994}%
  \BibitemOpen
  \bibfield  {author} {\bibinfo {author} {\bibfnamefont {E.~K.}\ \bibnamefont
  {Rodriguez}}, \bibinfo {author} {\bibfnamefont {A.}~\bibnamefont {Hoger}}, \
  and\ \bibinfo {author} {\bibfnamefont {A.~D.}\ \bibnamefont {McCulloch}},\
  }\bibfield  {title} {\enquote {\bibinfo {title} {Stress-dependent finite
  growth in soft elastic tissues},}\ }\href {\doibase
  10.1016/0021-9290(94)90021-3} {\bibfield  {journal} {\bibinfo  {journal}
  {Journal of Biomechanics}\ }\textbf {\bibinfo {volume} {27}},\ \bibinfo
  {pages} {455 -- 467} (\bibinfo {year} {1994})}\BibitemShut {NoStop}%
\bibitem [{\citenamefont {DiCarlo}\ and\ \citenamefont
  {Quiligotti}(2002)}]{DiCarlo:2002}%
  \BibitemOpen
  \bibfield  {author} {\bibinfo {author} {\bibfnamefont {A.}~\bibnamefont
  {DiCarlo}}\ and\ \bibinfo {author} {\bibfnamefont {S.}~\bibnamefont
  {Quiligotti}},\ }\bibfield  {title} {\enquote {\bibinfo {title} {Growth and
  balance},}\ }\href {\doibase 10.1016/S0093-6413(02)00297-5} {\bibfield
  {journal} {\bibinfo  {journal} {Mechanics Research Communications}\ }\textbf
  {\bibinfo {volume} {29}},\ \bibinfo {pages} {449--456} (\bibinfo {year}
  {2002})}\BibitemShut {NoStop}%
\bibitem [{\citenamefont {Gurtin}(2000)}]{Gurtin:2000}%
  \BibitemOpen
  \bibfield  {author} {\bibinfo {author} {\bibfnamefont {M.~E.}\ \bibnamefont
  {Gurtin}},\ }\href@noop {} {\emph {\bibinfo {title} {Configurational Forces
  as Basic Concepts of Continuum Physics}}}\ (\bibinfo  {publisher}
  {Springer},\ \bibinfo {year} {2000})\BibitemShut {NoStop}%
\bibitem [{\citenamefont {Nardinocchi}\ and\ \citenamefont
  {Teresi}(2016)}]{JAP:2016}%
  \BibitemOpen
  \bibfield  {author} {\bibinfo {author} {\bibfnamefont {P.}~\bibnamefont
  {Nardinocchi}}\ and\ \bibinfo {author} {\bibfnamefont {L.}~\bibnamefont
  {Teresi}},\ }\bibfield  {title} {\enquote {\bibinfo {title} {Actuation
  performances of anisotropic gels},}\ }\href {\doibase 10.1063/1.4969046}
  {\bibfield  {journal} {\bibinfo  {journal} {Journal of Applied Physics}\
  }\textbf {\bibinfo {volume} {120}},\ \bibinfo {pages} {215107} (\bibinfo
  {year} {2016})}\BibitemShut {NoStop}%
\bibitem [{Note1()}]{Note1}%
  \BibitemOpen
  \bibinfo {note} {See \cite {Curatolo:2017_M,Curatolo:2019} for further
  details.}\BibitemShut {Stop}%
\bibitem [{Note2()}]{Note2}%
  \BibitemOpen
  \bibinfo {note} {The Jacobians $J_a$ and $J$ have a key role in changing
  volume-elements. Precisely, let $dV_d$, $dV_a$, and $dv$ be the
  dry-reference, the remodeled, and the actual volume-elements, respectively.
  Then, they are related by the Jacobians as \begin {equation}\label {vols}
  dV_a=J_a\protect \tmspace +\thinmuskip {.1667em}dV_d\hskip 1em\relax \protect
  \textrm {and}\hskip 1em\relax dv =J\protect \tmspace +\thinmuskip
  {.1667em}dV_d = \protect \frac {J}{J_a}\protect \tmspace +\thinmuskip
  {.1667em}dV_a. \end {equation} In these terms, equation \protect \textup
  {\hbox {\mathsurround \z@ \protect \normalfont (\ignorespaces \ref
  {vc}\unskip \@@italiccorr )}} can be rewritten as \begin {equation}\label
  {sol} dv=dV_a+dV_{\protect \rm {sol}}\protect \tmspace +\thinmuskip
  {.1667em}, \end {equation} with $dV_{\protect \rm {sol}}=\Omega \protect
  \tmspace +\thinmuskip {.1667em}c\protect \tmspace +\thinmuskip {.1667em}dV_d$
  denoting the liquid volume-element. So, using these volume relationships,
  free energies per unit remodeled volume as $\varphi _e$ and $\varphi _m$ are
  transformed into free energies per unit dry volume as $\psi \protect \tmspace
  +\thinmuskip {.1667em}dV_d= (\varphi _e +\varphi _m)\protect \tmspace
  +\thinmuskip {.1667em}dV_a= J_a(\varphi _e +\varphi _m)\protect \tmspace
  +\thinmuskip {.1667em}dV_d$.}\BibitemShut {Stop}%
\bibitem [{Note3()}]{Note3}%
  \BibitemOpen
  \bibinfo {note} {An effective strain measures a change in length from the
  remodeled state.}\BibitemShut {Stop}%
\bibitem [{\citenamefont {Ideses}\ \emph {et~al.}(2018)\citenamefont {Ideses},
  \citenamefont {Erukhimovitch}, \citenamefont {Brand}, \citenamefont
  {Jourdain}, \citenamefont {Salmeron}, \citenamefont {Gabinet}, \citenamefont
  {Safran}, \citenamefont {Kruse},\ and\ \citenamefont
  {Bernheim-Groswasser}}]{Bernheim:2018_n}%
  \BibitemOpen
  \bibfield  {author} {\bibinfo {author} {\bibfnamefont {Y.}~\bibnamefont
  {Ideses}}, \bibinfo {author} {\bibfnamefont {V.}~\bibnamefont
  {Erukhimovitch}}, \bibinfo {author} {\bibfnamefont {R.}~\bibnamefont
  {Brand}}, \bibinfo {author} {\bibfnamefont {D.}~\bibnamefont {Jourdain}},
  \bibinfo {author} {\bibfnamefont {J.}~\bibnamefont {Salmeron}, \bibfnamefont
  {Hernandez}}, \bibinfo {author} {\bibfnamefont {U.}~\bibnamefont {Gabinet}},
  \bibinfo {author} {\bibfnamefont {S.}~\bibnamefont {Safran}}, \bibinfo
  {author} {\bibfnamefont {K.}~\bibnamefont {Kruse}}, \ and\ \bibinfo {author}
  {\bibfnamefont {A.}~\bibnamefont {Bernheim-Groswasser}},\ }\bibfield  {title}
  {\enquote {\bibinfo {title} {Spontaneous buckling of contractile poroelastic
  actomyosin sheets},}\ }\href@noop {} {\bibfield  {journal} {\bibinfo
  {journal} {Nature Communications}\ }\textbf {\bibinfo {volume} {9}},\
  \bibinfo {pages} {2461} (\bibinfo {year} {2018})}\BibitemShut {NoStop}%
\end{thebibliography}%

\end{document}